\documentclass[twoside,12pt]{article}
\usepackage{epsfig}

\newcommand{\be}{\begin{equation}}
\newcommand{\ee}{\end{equation}}
\newcommand{\bea}{\begin{eqnarray}}
\newcommand{\eea}{\end{eqnarray}}
\newcommand{\tha}{\theta_{12}}
\newcommand{\thb}{\theta_{13}}
\newcommand{\thc}{\theta_{23}}
\newcommand{\ea}{{\it et al.}}

\begin{document}

\title{ \vspace{1cm} Solar Neutrinos}
\author{R.G.H. Robertson\\
\\
Center for Experimental Nuclear Physics and Astrophysics, \\
and Department of Physics,
University of Washington, Seattle, WA 98195, USA.}
\maketitle
\centerline{\framebox{Dedicated to the memory of John Bahcall}}
\begin{abstract} Experimental work with solar neutrinos has illuminated the properties of neutrinos and tested models of how the sun produces its energy.  Three experiments continue to take data, and at least seven are in various stages of planning or construction.  In this review, the current experimental status is summarized, and future directions explored with a focus on the effects of a non-zero $\thb$  and the interesting possibility of directly testing the luminosity constraint.  Such a confrontation at the few-percent level would provide a prediction of the solar irradiance tens of thousands of years in the future for comparison with the present-day irradiance.  A model-independent analysis of existing low-energy data shows good agreement between the neutrino and electromagnetic luminosities at the $\pm 20$\% level.
\end{abstract}
%\eject
%\tableofcontents
\section{Introduction}
The first solar neutrino experiment, the Cl-Ar radiochemical detector \cite{ClAr} built by Ray Davis, Jr. and his colleagues, provided the only direct information about solar neutrinos in the seemingly endless interval between 1968 and 1988.  Towards the end of that period the Kamiokande proton-decay detector was being outfitted with new electronics in order to lower the threshold sufficiently to see the neutrinos from $^8$B decay in the sun, which it did in 1989 \cite{Hirata:1989zj}. That period also saw initiation by Vladimir Gavrin's group in the Soviet Union at the newly built Baksan Laboratory of a gallium radiochemical experiment, `SAGE',  to see if the $pp$ neutrinos were really there.  A similarly motivated effort in the US was abortive, and Till Kirsten launched the Gallex experiment with a mainly European team at Gran Sasso. The ``Standard Solar Model'' (SSM) constructed by Bahcall \cite{Bahc5} predicted that Ga experiments should see a neutrino capture rate of 136 SNU (captures per $10^{36}$ target atoms per second).  A rate above 69 SNU was thought to be indicative of errors in the SSM, while a smaller rate would require new neutrino physics.  That division, based on the rate in Ga of the $pp$ reaction by itself, today seems naive, and with much merriment Nature produced exactly 69 SNU in the Ga experiments.  SAGE reported its first results in 1990 \cite{Gavrin:1990wy,Abdurashitov:1994bc}, and Gallex in 1992 \cite{Anselmann:1993ct}.  Every experiment reported rates far below the expectations of the SSM.

The idea that neutrino physics might be responsible for the solar neutrino problem \cite{Gribov:1968kq} was initially greeted with skepticism.  Perhaps mixing might take place, but quark mixing angles were known to be small, and why should neutrinos be any different?  It was revived by the theoretical discovery of matter enhancement, the Mikheyev-Smirnov-Wolfenstein effect \cite{Mikheyev:1989dy,Wolfenstein:1977ue}, which allowed small vacuum mixing angles to be greatly amplified in matter.  (Once again, Nature laughed and gave us nearly maximal mixing {\em and} the MSW effect.)

In 1985 Herb Chen, George Ewan, and collaborators proposed the construction of a large, real-time heavy-water Cherenkov detector  to measure the $^8$B rate in both charged-current and neutral-current interactions, thus decoupling the neutrino physics from the astrophysics \cite{Chen:1985na,Aardsma:1987ph}.   The Sudbury Neutrino Observatory reported its first results in 2001 \cite{Ahmad:2001an}, comparing the pure CC rate to the elastic scattering (ES) rate seen in the Super-Kamiokande detector \cite{SK}.  The conclusion was that indeed the predicted neutrino flux was all there but that 2/3 of the electron neutrinos had converted to mu and tau neutrinos on their way to earth.  With steadily increasing precision, the measurements showed also the presence of neutrons liberated by neutral-current interactions with deuterium, again at 3 times the rate that would be expected from the CC data in the absence of neutrino oscillations.

The solar neutrino problem has been resolved, and neutrino flavor change demonstrated in an appearance experiment.  Neutrino oscillations and mass explain the observed effects well. The detailed model developed by Bahcall and many other astrophysicists was found to be astonishingly good, predicting the central temperature of the sun to a stunning 1\% accuracy.  The Standard Model of particles and fields must now be modified to include massive neutrinos. 

What does the future hold for solar neutrinos?  Only the radiochemical experiments provide information about the part of the spectrum from 0 to 5 MeV, where $>$ 99\% of the neutrinos reside.  To consider the consequences of assuming all is well there, one need only imagine Davis and Bahcall making the same assumption in 1965 and deciding to work on something else.  As it turns out, that region is where the transition from matter enhancement to vacuum oscillations takes place.  The detailed behavior of the transition is quite sensitive to the presence of nonstandard interactions, and a spectroscopic measurement of neutrinos there would be definitive.   Another interesting objective is to realize that, with sufficiently precise data, one could directly test the relationship between neutrino luminosity and radiant energy luminosity.   The fact that neutrinos from the center of the sun reach earth in 8 minutes, while photons take some 40,000 years \cite{Bahcall:1989ks}, would yield an eerie look at the future of the sun's life-giving energy. 

\section{Solar Neutrino Experiment Results}

Three experiments,  SAGE, Super-Kamiokande (SK), and SNO are currently in operation.  A fourth, KamLAND, is a terrestrial reactor antineutrino experiment, but it provides information intimately related to the solar neutrino data.   Taken together with the earlier results from Cl-Ar, Gallex, and GNO, the data from those experiments provide a remarkably precise picture of the mixing of two neutrinos and of the flux of high-energy neutrinos from the sun.

\subsection{SAGE}

\begin{figure}
\begin{center}
\begin{minipage}[t]{12 cm}
\epsfig{file=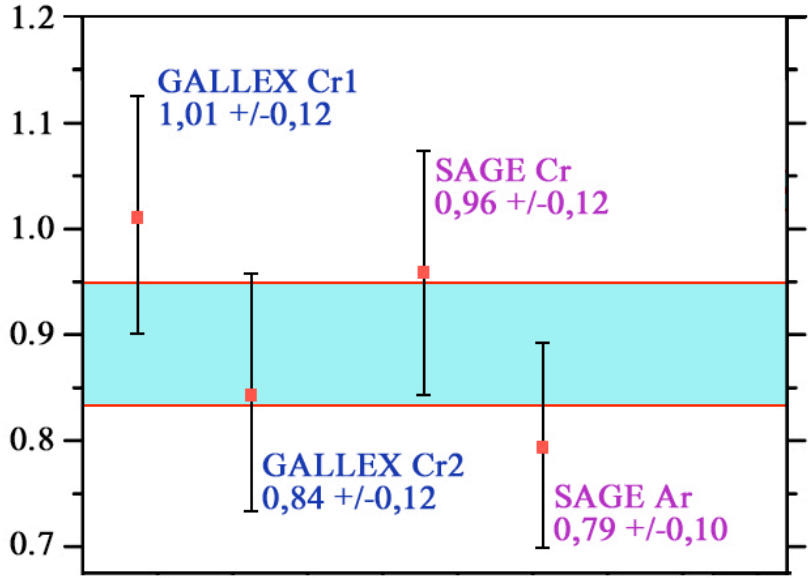,scale=0.5}
\end{minipage}
\begin{minipage}[t]{16.5 cm}
\caption{Measured neutrino capture rates in Gallex and SAGE with sources of $^{51}$Cr and $^{37}$Ar \protect\cite{SAGE2}. 
\label{fig:SAGE}}
\end{minipage}
\end{center}
\end{figure}

Beginning in January, 1990 with 30 Mg of Ga metal, SAGE \cite{SAGE1} has continuously recorded solar neutrinos via  the reaction, 
\begin{center}
 \begin{tabular}{ll}
        $ \nu_x + {^{71}{\rm Ga}} \rightarrow {^{71}{\rm Ge}} + e^- -0.233{\rm \ MeV}$  \hspace{0.5in} &  \\        
 \end{tabular}
\end{center}
In 1995 the target mass was increased to approximately 50 Mg, which, together with numerous improvements in the low-background proportional counters used to detect the decay of 11-day $^{71}$Ge,  has resulted in much improved statistical accuracy.  Recent work has focussed on the testing and calibration of the experiment with intense artificially produced antineutrino sources of $^{51}$Cr and $^{37}$Ar \cite{SAGE3}.    The feasibility of preparing an intense, pure source of $^{37}$Ar via $^{40}$Ca(n,$\alpha$){$^{37}$Ar} in a fast reactor as originally suggested by Haxton \cite{Haxton37} has been convincingly demonstrated.  The results, however, are a little surprising (Fig. \ref{fig:SAGE}) and show a consistently lower rate in SAGE and Gallex for both $^{51}$Cr and $^{37}$Ar sources than is expected from the efficiencies and cross sections in use.  The SAGE collaboration's conclusion \cite{SAGE2} is, ``The source experiments with Ga should be considered to be a determination of the neutrino capture cross section.''   Since the ground-state cross section is fixed from $\beta$ decay by detailed balance, the correction falls entirely on excited-state cross sections, which already play a relatively minor role in low-energy neutrino capture, and is therefore substantial. 

\subsection{Kamiokande and Super-Kamiokande}

Built as  proton-decay experiments, SK and its predecessor Kamiokande have produced a remarkable view of the sun through detection of the elastic scattering of $^8$B neutrinos from electrons: 
\begin{center}
 \begin{tabular}{ll}
        $ \nu_x + e^- \rightarrow \nu_x + e^-$ \hspace{0.5in} & (ES)\\        
 \end{tabular}
\end{center}
SK is very large, with a fiducial mass of 22.5 Gg. The collaboration has recently released a detailed paper \cite{SK2} on the 1496 days of solar neutrino data collected from April 1996 through July 2001.  The experiment is at this writing shut down while the number of photomultipliers is restored to the original 11,000 it had before the Nov. 12, 2001 accident.  It is anticipated that SK will return to operation with full coverage in the spring of 2006. 

\subsection{SNO}

Heavy water (D$_2$O) permits three distinct reactions with solar neutrinos, 
\begin{center}
 \begin{tabular}{ll}
    $\nu_e + d \rightarrow p + p + e^- -1.44{\rm \ MeV}$\hspace{0.5in} & (CC)\\
    $ \nu_x + d \rightarrow p + n + \nu_x -2.22{\rm \ MeV} $ & (NC)\\
    $ \nu_x + e^- \rightarrow \nu_x + e^-$  & (ES)\\        
 \end{tabular}
\end{center}
Beginning in November, 1999, the SNO experiment \cite{Boger:1999bb} has run in three configurations, pure D$_2$O \cite{SNO3}, D$_2$O with the addition of 0.195\% by weight of NaCl \cite{SNO4}, and D$_2$O with the deployment of an array of $^3$He-filled proportional counters.  SNO has used a fiducial mass of 770 Mg. The last phase is still in progress and will be completed December 31, 2006, at which point the heavy water will be returned to its owners.  

\subsection{KamLAND}

Built in the site originally occupied by Kamiokande I, KamLAND is a liquid-scintillator detector sensitive to antineutrinos from reactors in Japan, which fortuitously happen to be situated in a roughly circular pattern about 185 km in radius centered on Kamioka.  The signal is very clean and free of most backgrounds thanks to the delayed coincidence between inverse beta decay and neutron capture:
\begin{center}
 \begin{tabular}{ll}
        $ \overline{\nu_e} + p \rightarrow n + e^+ -1.804{\rm \ MeV}$ \hspace{0.5in} & (Inverse beta)\\  
	 $ n + p \rightarrow d + \gamma +2.223 {\rm \ MeV} $ & (n capture)\\        
 \end{tabular}
\end{center}
One background, ($\alpha$,n) induced by radon progeny $^{210}$Po, evades the strategy and required separate calculation and subtraction.	 The fiducial mass of KamLAND was initially chosen to be 408 Mg \cite{KL1}, and in a more recent analysis \cite{KL2} of the complete data set from March, 2002 to January, 2004,  543.7 Mg.  The baseline and energy spectrum of the reactor antineutrinos give KamLAND the ability to make a pinpoint determination of the mass splitting $\Delta m_{12}^2$, whereas  the solar experiments, both because of matter enhancement and the direct measurement by SNO of the CC/NC ratio, excel at determining the mixing angle $\tha$.  
	 
\subsection{Cl-Ar}

The historic experiment of Davis \cite{ClAr}, which recorded the reaction
\begin{center}
 \begin{tabular}{ll}
        $ \nu_e + {^{37}{\rm Cl}} \rightarrow {^{37}{\rm Ar}} + e^- -0.814{\rm \ MeV}$  \hspace{0.5in} &  \\        
 \end{tabular}
\end{center}
using 615 Mg of C$_2$Cl$_4$, took its last data in 1998, but the results remain very important in modern analyses.  The radiochemical experiments are integral and so do not by themselves give spectroscopic information, but in combination with other experiments their differing thresholds do yield a coarse spectroscopy.  The Cl-Ar and Ga results together map out the transition from matter-enhanced to vacuum oscillations.
 
 \subsection{Gallex and GNO}
 
 The Gallex experiment in Gran Sasso ran with 30 Mg of Ga in the form of GaCl$_3$ in the period May, 1991 to January, 1997 \cite{Hampel:1998xg}.  That Ga was inherited by the successor experiment GNO \cite{Altmann:2005ix}, which  ran successfully from May, 1998 to April, 2003, when it became a casualty of an accidental release of pseudocumene into the environment at Gran Sasso.   It is worth emphasizing how valuable it was for the scientific community to see the highly consistent results from Gallex and SAGE, obtained with very different technical approaches by spirited collaborations that would as soon have seen their competition shown to be in error.  For many in the larger fields of particle and nuclear physics, this marked a turning point at which the possibility of new neutrino physics had to be taken seriously.
 
 \subsection{Results}
 
Table \ref{tab:results} gathers in one place the results of solar neutrino experiments, KamLAND, and satellite measurements of the electromagnetic solar irradiance. 
\begin{table}
\caption{Results from solar neutrino experiments, KamLAND, and the solar irradiance.}
\medskip
\begin{center}
\begin{tabular}{lrrrlc} 
\hline
\hline
Measurement & value & stat.  &  syst.  & units  & Ref.  \\
\hline
SNO CC/NC & 0.340 & 0.023 & $^{+0.029}_{-0.031}$ & & \cite{SNO4}  \\
SNO NC+CC & 4186	 & 	65	& 244 & Events & \cite{SNO4} \\
SNO ES flux as if $\nu_e$ & 2.35 & 0.22 & 0.15 & $10^{6}$ cm$^{-2}$ s$^{-1}$ & \cite{SNO4} \\
SK ES flux as if $\nu_e$ & 2.35 & 0.02 & 0.08 & $10^{6}$ cm$^{-2}$ s$^{-1}$ & \cite{SK2} \\
Chlorine & 2.56 &	0.16	& 0.14 & SNU & \cite{ClAr} \\
SAGE & 67.2	& 3.7 & $^{+3.6}_{ -3.2}$ & SNU & \cite{SAGE2} \\
Gallex/GNO & 69.3	& \multicolumn{2}{c}{5.5}  & SNU & \cite{Altmann:2005ix} \\
KamLAND $P_{ee}$ & 0.658 &	0.044 & 	0.047 &  & \cite{KL2} \\
Solar Irradiance & 85.31 & &  0.34 & $10^{10}$ MeV cm$^{-2}$ s$^{-1}$  & \cite{BP} \\
\hline
\hline
\end{tabular}
\end{center}
\label{tab:results}
\end{table}

\section{Physics from Solar and Reactor Neutrinos}

With the quite extensive and precise data now available,  detailed analyses of neutrino physics and solar astrophysics can be made.  Most analyses of neutrino oscillations, until recently, have been in the context of 2-neutrino mixing because the solar and atmospheric physics separate fairly cleanly.  

\subsection{Two-mass Mixing}

\begin{figure}[tb]
\begin{center}
\begin{minipage}[t]{8 cm}
\epsfig{file=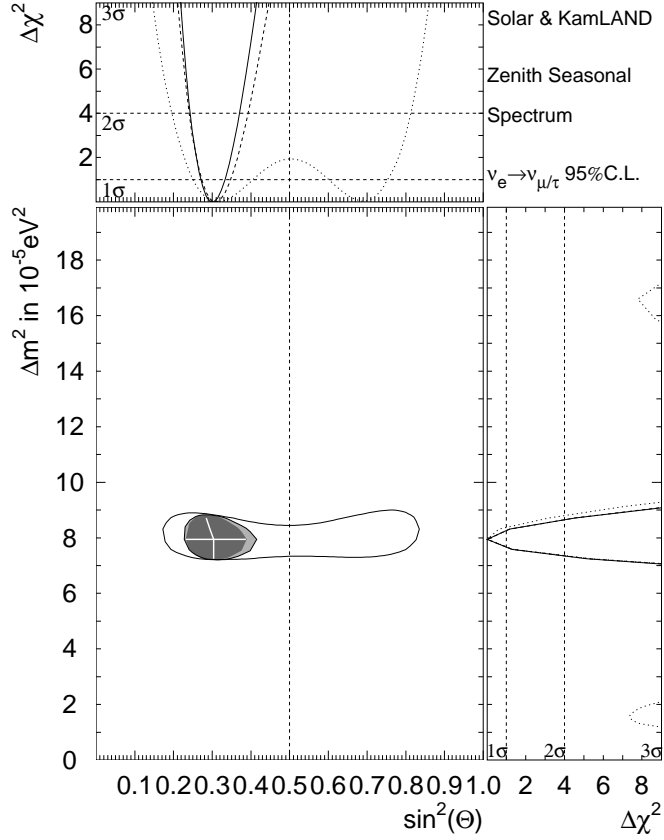,scale=1}
\end{minipage}
\begin{minipage}[t]{16.5 cm}
\caption{The 95\% confidence-level contours for three scenarios: (outermost) KamLAND alone, (middle) KamLAND plus SNO and SK, and (innermost), also including other solar neutrino data and the SSM \protect\cite{SK2}. 
\label{fig:SKcontour}}
\end{minipage}
\end{center}
\end{figure}

In the context of two active mass eigenstates, a global analysis of solar and reactor data yields \cite{SNO4}  for the joint 2-dimensional 1-$\sigma$ boundary,  
\begin{eqnarray*}
\Delta m^2 & = & 8.0^{+0.6}_{-0.4}\times 10^{-5} {\rm \ eV}^2 \\
\theta &= & 33.9^{+2.4}_{-2.2} {\rm \ degrees} 
\end{eqnarray*}
%\medskip
For the marginalized 1-$\sigma$ uncertainties, the results are:
\begin{eqnarray*}
\Delta m^2 & = & 8.0^{+0.4}_{-0.3}\times 10^{-5} {\rm \ eV}^2 \\
\theta &= & 33.9^{+1.6}_{-1.6} {\rm \ degrees} 
\end{eqnarray*}
Mixing in the solar sector is certainly large, but at the same time maximal mixing is ruled out at more than $5\sigma$.  There is residual model dependence in these results arising from the use of SSM fluxes for {\em pp}, {\em pep}, $^7$Be, CNO, and {\em hep} neutrinos.   It is, however, quite small as can be seen in Fig. \ref{fig:SKcontour}, from \cite{SK2}.  The difference between the innermost and middle contours is from the inclusion of data interpreted via the SSM.  The $^8$B flux is allowed to float throughout.

\subsection{Three-mass Mixing}

The electron neutrino in principle contains components of all 3 mass eigenstates:
\begin{eqnarray*}
U_{e1} & = & \cos\theta_{12}\cos\theta_{13} \\
U_{e2} & = & \sin\theta_{12}\cos\theta_{13} \\
U_{e3} & = & \sin\theta_{13} e^{-i\delta}  
\end{eqnarray*}
but the small size of $\thb$, known to be less than 10 degrees, leads to the convenient and well-known simplification to two-mass mixing with a sacrifice in precision that is for most applications minor.   However, solar neutrino and reactor experiments are reaching precisions of a few percent and it is germane to ask, 
\begin{itemize}
\item Do solar neutrinos have anything to contribute to the determination of $\thb$, and,
\item Does inclusion of the third state and the attendant uncertainty in $\thb$ affect the determinations of the solar mixing parameters?
\end{itemize}

The large mass gap for 13 mixing and the low energies of solar neutrinos means that any effects in the sun from it are independent of energy.  The matter resonance would occur at 190 MeV at the center of the sun.   On this basis, Fogli {\em et al.} \cite{fogli2} give a relationship between the 3- and 2-neutrino scenarios, 
%...............................................................................
\begin{equation}
P_{ee}^{3\nu}(\delta m^2,\tha,\thb) = \sin^4\thb +
\cos^4\thb\cdot P_{ee}^{2\nu}(\delta m^2,\tha)
\Big|_{N_e\to \cos^2\thb \,N_e}\ .
\label{e2}
\end{equation}
%where $\phi = \theta_{13}$ and $\omega = \theta_{12}$.

Qualitatively this consists of an energy-independent conversion of some $\nu_e$ flux into $\nu_2$ and $\nu_3$ via $\thb$, and a dilution factor of $\cos^2\thb$  included with the electron density since the beam is no longer prepared as exactly $\nu_e$ when it crosses the 12 matter resonance.

\begin{figure}[tb]
\begin{center}
\begin{minipage}[]{16.5 cm}
\epsfig{file=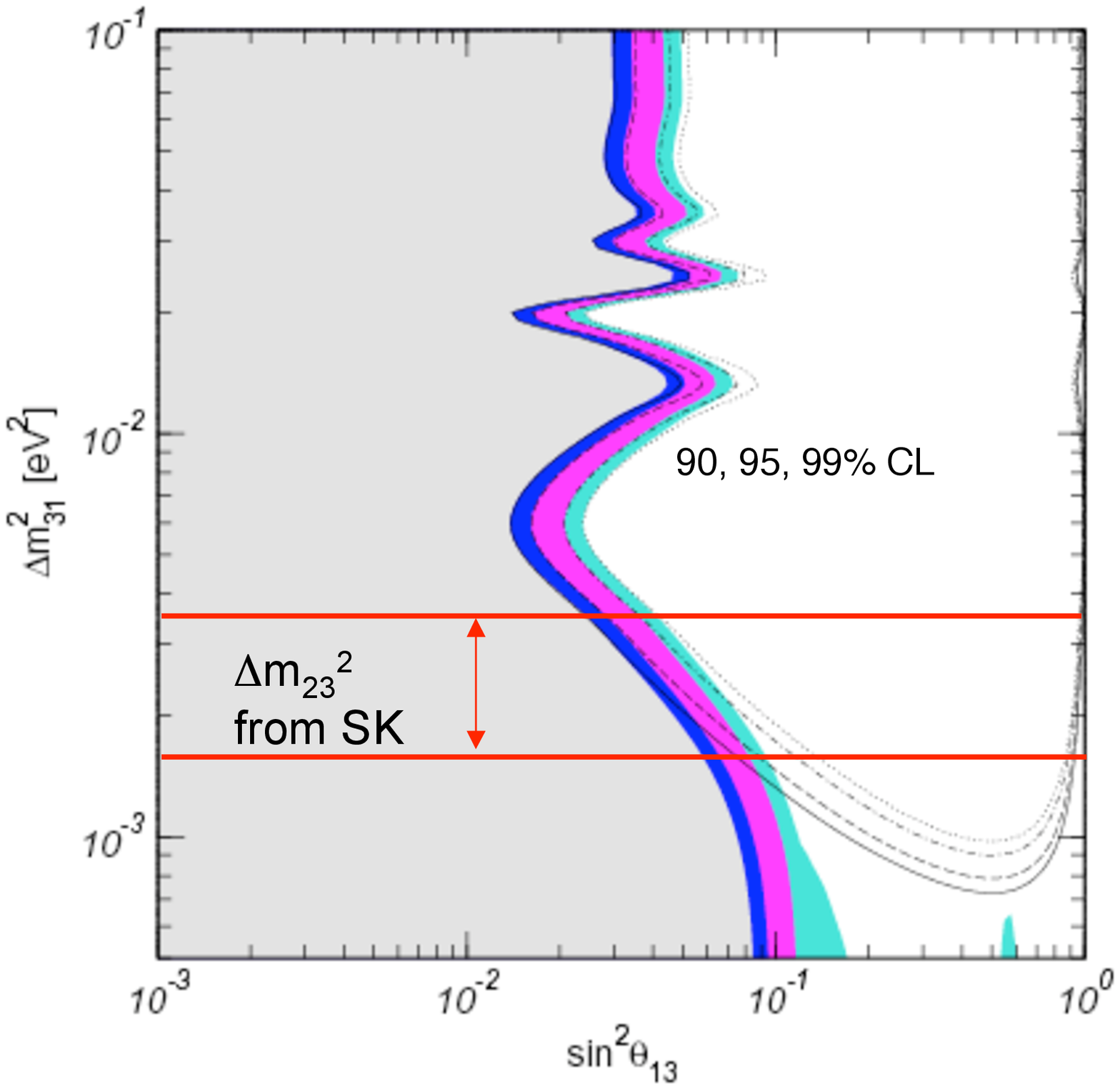,scale=.45}
\epsfig{file=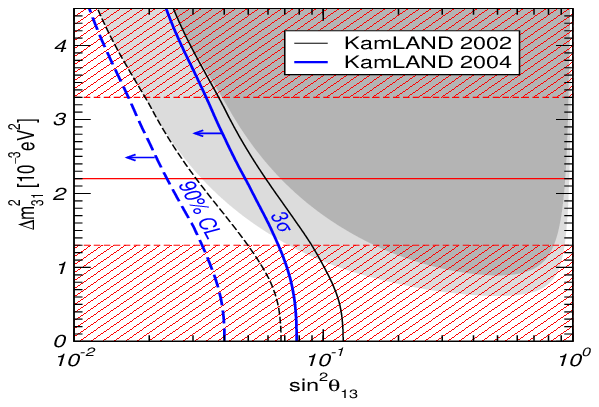,scale=1.6}
\end{minipage}
\begin{minipage}[t]{16.5 cm}
\caption{(left) Limits on $\thb$ from Chooz (lines, 90\%, 95\%, 99\%, and 3$\sigma$), and from Chooz+solar+KamLAND (colored regions) \protect\cite{maltoni}.  (right) Limits updated for new data \protect\cite{Valle}.
\label{fig:maltoni8}}
\end{minipage}
\end{center}
\end{figure}

Atmospheric neutrinos give, again in a 2-mass mixing description, $\theta_{23} = 45 \pm 9$ degrees and $\Delta m_{23}^2 = 2.1_{-0.6}^{+1.3} \times 10^{-3}$ eV$^2$ (90\% CL) \cite{SK3}.  A global analysis of all available data by Maltoni {\em et al.} \cite{maltoni}  in 2003 summarized the situation and drew attention to the role of solar and KamLAND reactor neutrino data in limiting  $\thb$.  Figure \ref{fig:maltoni8} is from that work, superimposed with modern limits \cite{SK3} on $\Delta m_{23}^2$.   One can see that near the low end of the mass range the tightest limits on $\thb$ were already coming from solar neutrinos and KamLAND.  The relationship between these experiments and $\thb$ began to be explored even before results were available from KamLAND \cite{concha}.  The right-hand panel is from a recent analysis \cite{Valle}  including the latest KamLAND data \cite{KL2}.

\begin{figure}[tb]
\begin{center}
\begin{minipage}[]{16.5 cm}
\epsfig{file=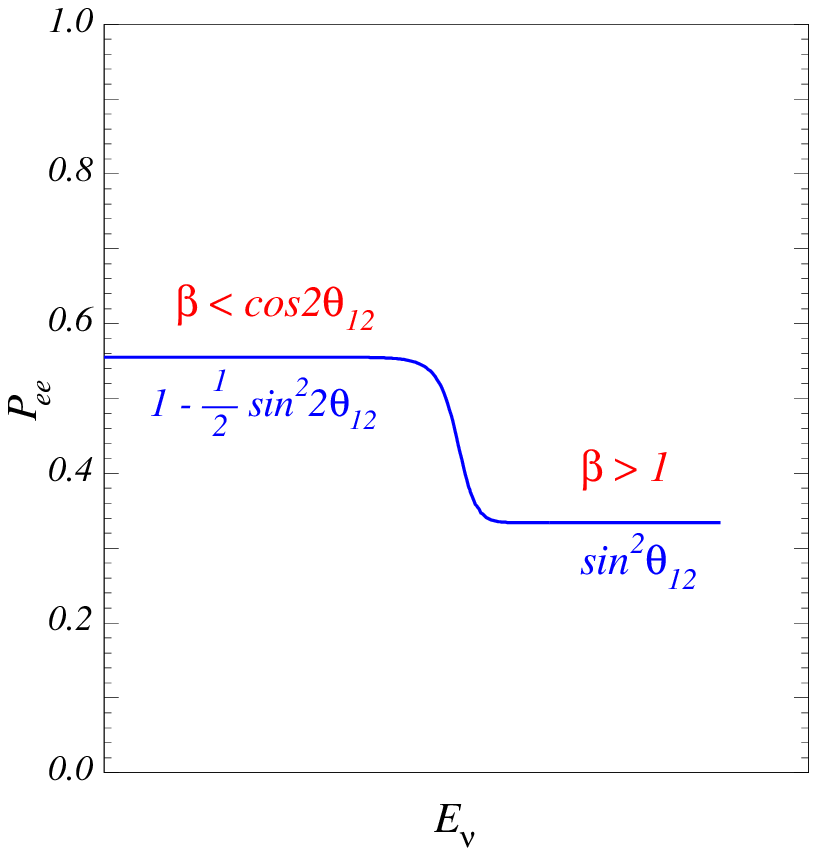,scale=.9}
\epsfig{file=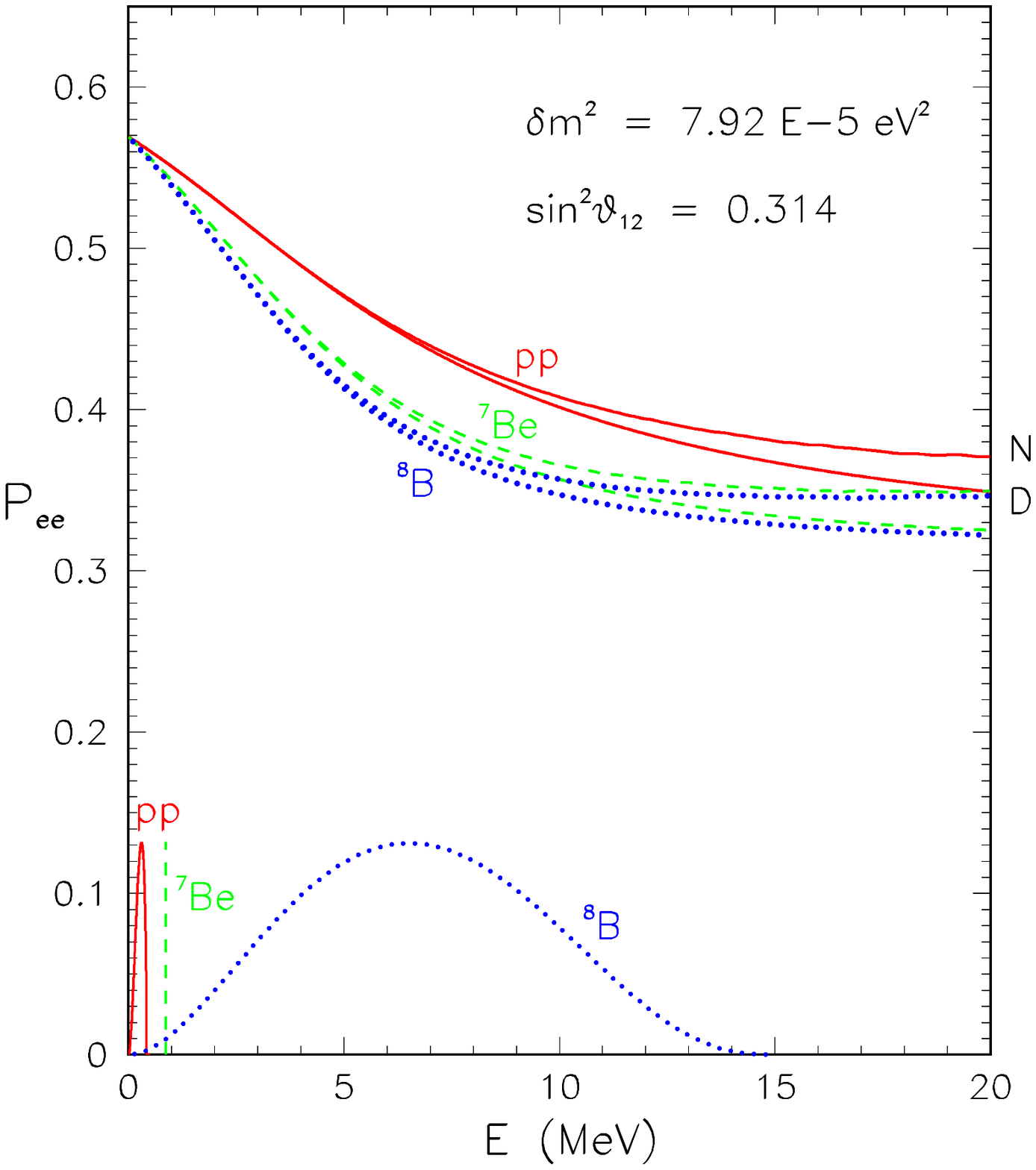,scale=.5}
\end{minipage}
\begin{minipage}[t]{16.5 cm}
\caption{Matter enhancement (MSW) in the LMA region. (left) Schematic, from \protect\cite{roadmap}; (right) Actual from \protect\cite{fogli3}, with $\thb=0$. 
\label{fig:matter}}
\end{minipage}
\end{center}
\end{figure}

Solar and reactor neutrinos allow a separation of the 12 and 13 effects by virtue of matter enhancement, which singles out the 12 component in the solar data.  The transition from no matter enhancement at low energies  to full matter enhancement above 5 MeV changes $P_{ee}$ from $1- \frac{1}{2}\sin^22\theta_{12}$ to $\sin^2\theta_{12}$.  This is illustrated schematically in \cite{roadmap}, from which the left-hand panel of Fig. \ref{fig:matter} is extracted.
The NC and ES measurements then define the total active flux, which, when compared to CC data, gives a ratio that depends on both $\theta_{12}$ and $\theta_{13}$ in the high-energy regime.  The final input comes from the total active flux normalization at low energies, which is derived mainly from the luminosity constraint and the standard solar model.  A comprehensive global analysis of all available oscillation data as of August, 2005, by the Bari group \cite{fogli3} gives the following results at the 95\% CL:
\begin{eqnarray*}
\sin^2\thb &=& 0.9_{-0.9}^{+2.3} \times 10^{-2} \\
\Delta m_{12}^2 &=& 7.92(1\pm0.09) \times 10^{-5} {\rm \ eV}^2, \\
\sin^2\tha &=& 0.314(1^{+0.18}_{-0.18}), \\
\Delta m_{23}^2 &=& 2.4(1^{+0.21}_{-0.26}) \times 10^{-3} {\rm \ eV}^2, \\
\sin^2\thc &=& 0.44(1^{+0.41}_{-0.22}), 
\end{eqnarray*}
The determination of the oscillation parameters can be made in an essentially model-independent way, although in most global analyses, such as \cite{fogli3}, it is customary to include the low-energy solar neutrino data by calculating $P_{ee}$ against the standard solar model \cite{BS05}.  As has been emphasized above (Fig. \ref{fig:SKcontour}), however, the low-energy solar neutrino data now play a very minor role in determination of the oscillation parameters, those parameters being fixed by SNO, SK, and KamLAND. This raises the interesting possibility of using that `free energy' in the low-energy data to make new tests of other physics.  In particular, a model-independent determination of the total solar neutrino luminosity can now be made and compared to the electromagnetic luminosity.

Global fits can be made numerically, but it is enlightening to look at a set of coupled equations that can be solved straightforwardly to provide best-fit parameters, uncertainties, and correlation coefficients. While not exact, the set of equations makes clear the interrelationships between various kinds of experiment and the parameters that can be determined from them. 

The simplifying assumptions include the following:
\begin{itemize}
\item The neutrino phenomenology is the LMA solution with 3 active neutrinos, and KamLAND plus SNO fix the mass to the so-called LMA-I solution.  This has the advantage of decoupling $\Delta m_{12}^2 $ from the calculations, since in this part of the LMA space $P_{ee}$ is independent of $\Delta m_{12}^2 $.
\item  The possibilities of sterile neutrino admixtures, non-standard interactions, and  violation of CPT are neglected.
\item The spectral distortions in both SK and SNO are negligible, connecting the fluxes measured in the experiments (above the 5 MeV threshold) to the total fluxes.
\item The  pp, pep, and $^7$Be neutrinos are in the vacuum oscillation region.  The `critical energies' in the sun for $^8$B, pp, and $^7$Be  are 1.8, 2.2, and 3.3 MeV respectively  \cite{roadmap}.  See Fig. \ref{fig:matter}.
\item The analysis considers only the solar and KamLAND inputs.  Chooz and atmospheric neutrinos also provide constraints on $\theta_{13}$.
\item The CNO flux is set to the SSM \cite{BS05} value, 0.8\%.  The flux is included with $\phi_7$, which then is not strictly a ``$^7$Be'' flux.
\end{itemize}

With this framework there are up to 5 unknowns, the 3 (total active) fluxes $\phi_1$, $\phi_7$, and $\phi_8$, and two mixing angles, $\theta_{12}$ and $\theta_{13}$.  The mass-squared difference $\Delta m_{12}^2 $ is fixed by  the KamLAND reactor oscillation experiment, and $\Delta m_{23}^2 $ by the atmospheric neutrino data.  Since the latter is much larger than the former,  $\Delta m_{13}^2 \simeq  \Delta m_{23}^2 $ for either hierarchy. To extract these 5 unknowns there are 7 equations relating them to experimental observables.

\begin{eqnarray}
P_{ee}^{SNO} & = & \epsilon_{\rm LMA}(\sin^2\tha \cos^4\thb + \sin^4\thb) \\
R_{\rm tot}^{SNO} & = &  \phi_8\left[\sigma_{d8}^{\rm N} + \sigma_{d8}^{\rm C}  \epsilon_{\rm LMA}(\sin^2\tha \cos^4\thb + \sin^4\thb)\right]  \\
\Phi_{\rm ES}^{SK, SNO} & = &   \eta^{-1} \phi_8 +  (1 - \eta^{-1}) \phi_e \nonumber\\
& = & \phi_8 \left[\eta^{-1} + (1-\eta^{-1}) \epsilon_{\rm LMA}(\sin^2\tha\cos^4\thb + \sin^4\thb) \right] \\
R^{\rm Cl} & = & \sigma_{C8}  \epsilon_{\rm LMA}\phi_8 (\sin^2\tha \cos^4\thb + \sin^4\thb) \nonumber \\
&& +( \sigma_{C1}\phi_1+ \sigma_{C7}\phi_7)\left[(1-\frac{1}{2}\sin^22\tha)\cos^4\thb + \sin^4\thb\right] \\
%&& + \sigma_{C7}\phi_7\left[(1-\frac{1}{2}\sin^22\tha)\cos^4\thb + \sin^4\thb\right] \\
R^{\rm Ga} & = & \sigma_{G8} \epsilon_{\rm LMA} \phi_8 (\sin^2\tha \cos^4\thb + \sin^4\thb) \nonumber \\
&& +(\sigma_{G1}\phi_1+ \sigma_{C7}\phi_7)\left[(1-\frac{1}{2}\sin^22\tha)\cos^4\thb + \sin^4\thb\right]  \\
%&& + \sigma_{G7}\phi_7\left[(1-\frac{1}{2}\sin^22\tha)\cos^4\thb + \sin^4\thb\right] \\
\frac{2I}{Q} & = & 0.980(1-0.088f_{pep})\phi_1+0.939(1-0.003f_{CNO})\phi_7+0.498\phi_8 \\
P_{ee}^{\rm KL} & = & \left[1-\sin^22\tha\overline{\sin^2\frac{\Delta m_{12}^2 L}{4E}}\right]\cos^4\thb +\sin^4\thb 
\end{eqnarray}

The first and second equations describe the electron neutrino survival probability (CC/NC ratio) and the total rate of NC+CC interactions in SNO.  This particular choice of representation of the NC and CC rates has the advantage of minimizing the correlation between the two equations.    The third equation is the equivalent electron neutrino flux measured by SK and SNO, where $\eta = 6.383$ is the cross section ratio for electron neutrinos relative to $\mu$ and $\tau$ neutrinos above 5 MeV.  The fourth and fifth equations give the rates in the Cl and Ga detectors in terms of the three (total active) flux components $\phi_1$, $\phi_7$, and $\phi_8$ and the cross sections $\sigma_i$.  The $\phi_1$ spectrum includes both the {\em pp} continuum and {\em pep} line features. The sixth equation is the luminosity constraint (see, for example, \cite{spiro,heeger,jnblum}).  The seventh equation is the (anti) neutrino survival probability in KamLAND, for which the effective distance argument is averaged over the various reactors that contribute to the signal.  From the best-fit parameters for the KL data alone, one finds  $$\overline{\sin^2\frac{\Delta m_{12}^2 L}{4E}} = 0.389.$$  A near-unity correction parameter $\epsilon_{\rm LMA}$ is introduced to correct for the small difference between the CC/NC ratio given by the simplified expression $\sin^2\tha \cos^4\thb + \sin^4\thb$ and the value measured and fitted in detailed numerical analyses such as that of Fogli {\ea} \cite{fogli3}.  The value found for $\epsilon_{\rm LMA}$ is 1.10.

In Table \ref{coeffs} the cross sections used are listed.  The value of $f_{pep}$ is $f_{pep} = 0.23(2)$\% and $f_{CNO} = 0.8\%$.  The cross-section uncertainties in the radiochemical experiments are propagated through the flux equations to be added in quadrature with the experimental uncertainties in the rate.

\begin{table}
\caption{Cross-section coefficients.}
\medskip
\begin{center}
\begin{tabular}{lllr} 
\hline
\hline
& 	& (Effective) &  \\
&  	& Cross Section	&  Reference \\
& & 10$^{-46}$ cm$^2$ &  \\
\hline
SNO & $\sigma_{D8}^{N}$ & 2630  &    \cite{SNO4} \\
	& $\sigma_{D8}^{C}$ & 8000 &    \cite{SNO4} \\
\hline
& $\sigma_{C1}$ & 16 $f_{pep}$  &    \protect{\cite{BP,BU}} \\
Cl-Ar & $\sigma_{C7}$ & 2.38(1 + 2.60$f_{CNO}$)  &    \protect{\cite{Bahc5}} \\
& $\sigma_{C8}$ & 11100  &   \protect{\cite{Trinder,Bahc6}} \\
\hline
& $\sigma_{G1}$ & 11.8(1 + 17$f_{pep}$)  &    \protect{\cite{BU,Hampel}} \\
Gallium & $\sigma_{G7}$ & 76.5(1 + 1.42$f_{CNO}$)  &    \protect{\cite{Haxton}}
\\ & $\sigma_{G8}$ & 24600  &    \protect{\cite{Bahc6}} \\
\hline
\hline
\end{tabular}
\end{center}
\label{coeffs}
\end{table}

\begin{figure}[]
\begin{center}
\begin{minipage}[]{18 cm}
\epsfig{file=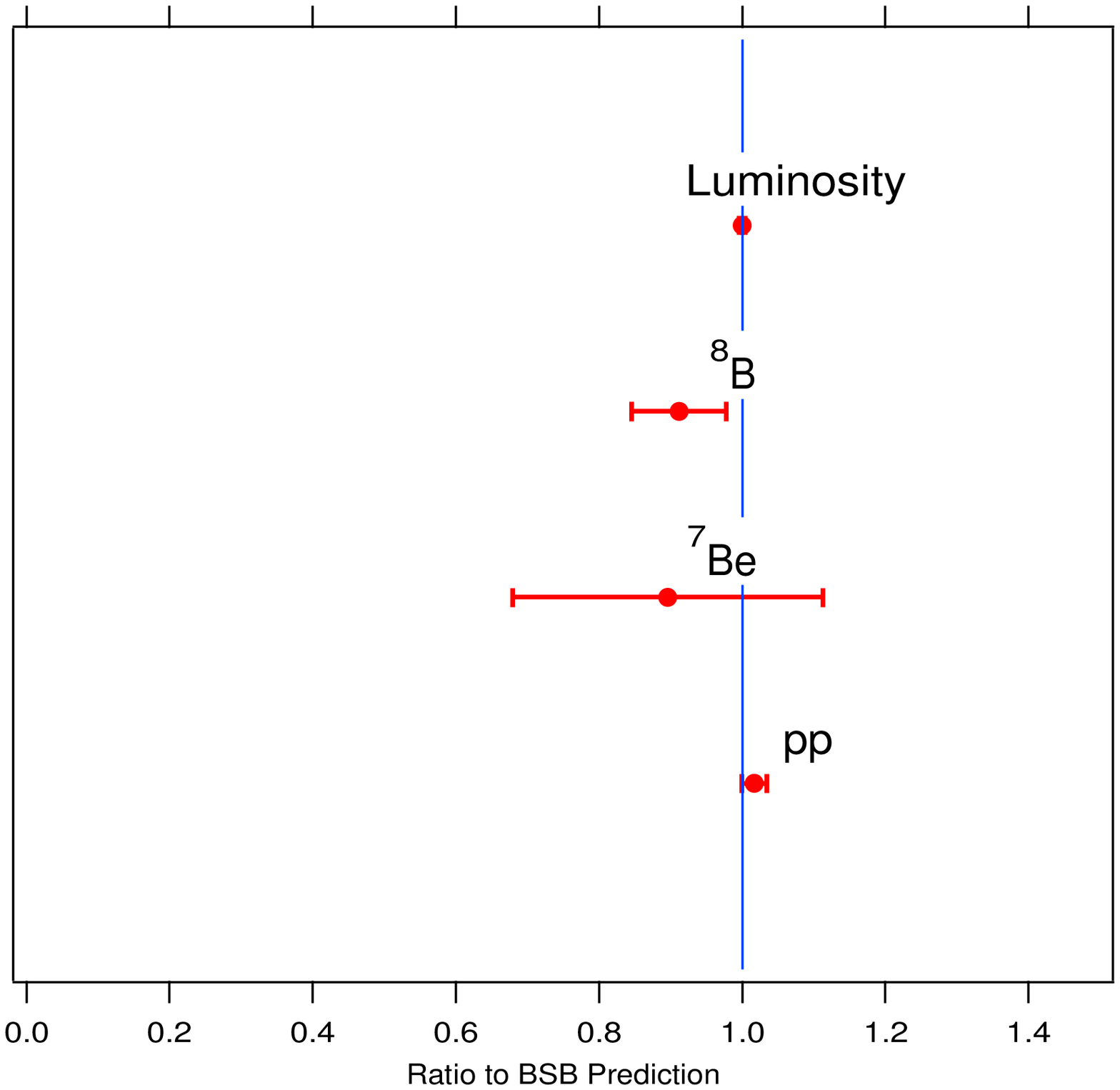,scale=.45}\epsfig{file=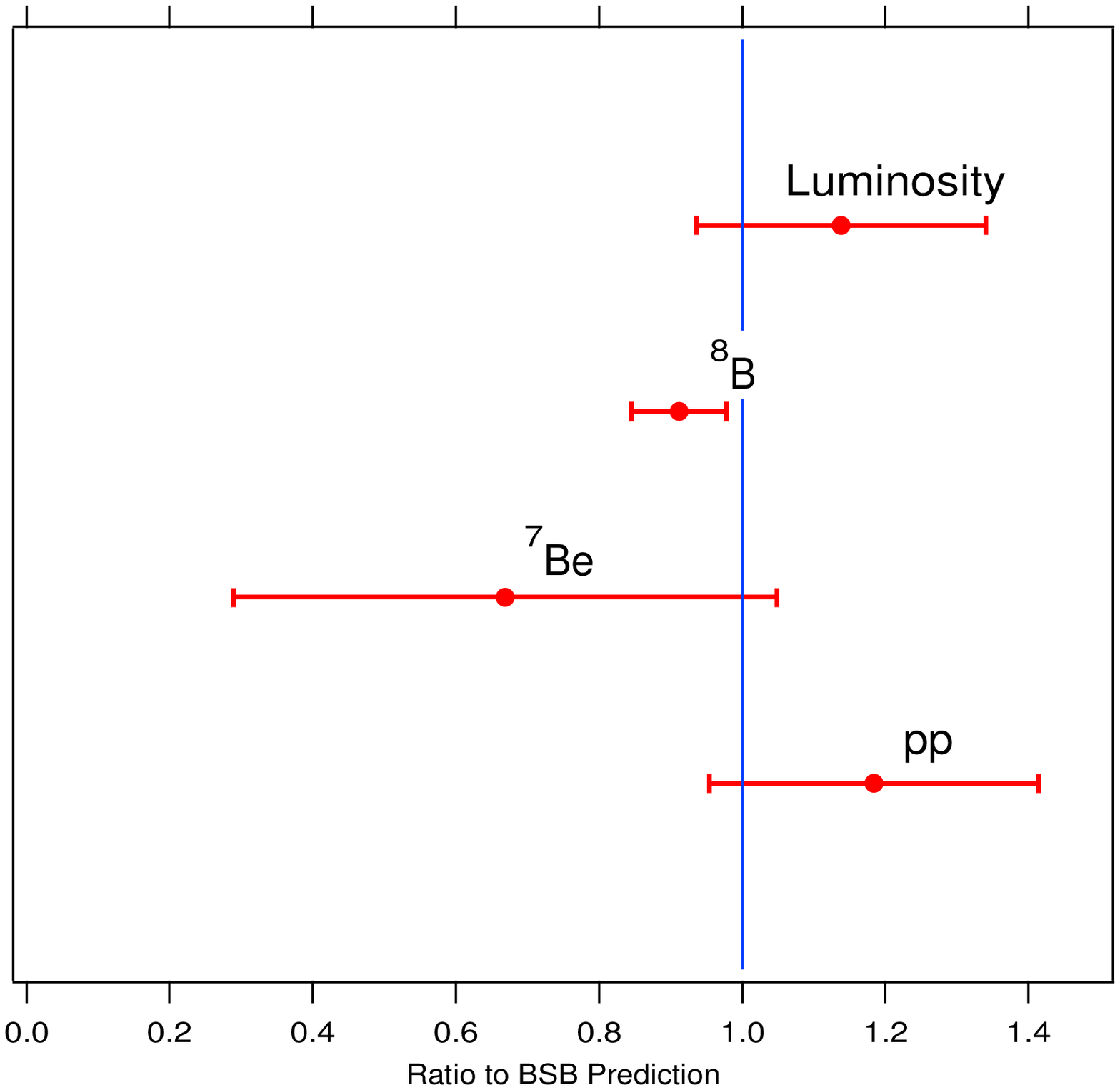, scale=0.45}
\end{minipage}
\begin{minipage}[t]{16.5 cm}
\caption{Model-independent determination of the low-energy fluxes and the solar luminosity by solar neutrino experiments and KamLAND.  The results are expressed as ratios with 1-$\sigma$ uncertainties to the BSB (OP) solar model \protect\cite{BS05} values.  The angles $\tha$ and $\thb$ were fixed at 34.1$^o$ and 5.4$^o$, respectively \cite{fogli3}.  The $^8$B flux depends significantly on the mixing angles, and the value and uncertainty shown are from the numerical fit of Fogli \ea \ \protect\cite{fogli3}. The left panel shows the results with the luminosity constraint, and the right without it.  (Interestingly, there is no indication of a significant problem with the low-energy fluxes in either case.)  This procedure sets a general and quite restrictive limit on the contribution of sterile neutrinos to the solar neutrino flux.
\label{fig:lum}}
\end{minipage}
\end{center}
\end{figure}

The experimentally determined rates and ratios used as input are listed in Table \ref{tab:results}, and the results are summarized in Fig. \ref{fig:lum}.   The uncertainties are large and will remain so until a determination of the $^7$Be flux is made, but it is remarkable how a model-independent analysis of the low-energy solar data together with a model-independent determination of neutrino oscillation parameters together produce  results very consistent with solar models.

%\end{document}

\section{Nuclear Astrophysics and Solar Neutrinos}

The SSM relies on a large body of painstaking laboratory work in a number of different research areas. In addition, there are laboratory inputs that affect directly the extraction of neutrino oscillation parameters in model-independent analyses.  The past few years have seen great progress in improving the accuracy with which important nuclear-physics inputs are known.  Much remains to be done, however.   For a comprehensive general summary, the reader is referred to the recent paper by Haxton, Parker, and Rolfs \cite{Haxton:2005rw}.   A number of additional reactions are included in the discussion  below.

\subsection{$^3$He($^3$He,2p)$^4$He and $^{14}$N(p,$\gamma$)$^{15}$O}

The commissioning of the LUNA accelerator in the Gran Sasso Laboratory has made possible two important measurements that would have been overwhelmed at the lowest energies by backgrounds if done on the surface.   The $^3$He($^3$He,2p)$^4$He reaction has been measured down to energies  completely inclusive of the Gamow peak where stellar burning takes place \cite{LUNA}.  The rate-determining step in the CNO cycle is the  $^{14}$N(p,$\gamma$)$^{15}$O reaction, and a precise determination of its rate was identified as a priority in the evaluation conducted in 1998 \cite{Adelberger:1998qm}.  That was accomplished both at LUNA \cite{Imbriani:2005jz} and at LENA \cite{LENA}, resulting in a recommended S-factor, 1.67 keV b, a factor of 2 smaller than the previously accepted value.   

%\subsection{CNO Neutrinos}

There is insufficient data for a completely model-independent analysis of the CNO flux from the sun, but in 2002, Bahcall, Gonzalez-Garcia, and Pe\~{n}a-Garay \cite{Bahcall:2002jt} carried out an analysis similar to that reported in this paper, a fit to the fluxes as free parameters under the assumption of the LMA solution, but invoking also the luminosity constraint.  The higher cross section for CNO neutrino capture as compared to {\em pp} and $^7$Be neutrinos then allows a limit to be set on the CNO flux.  Bahcall \ea \ concluded that the solar CNO luminosity was (at 3$\sigma$) less than 7.3\% of the total luminosity.

Before the new measurements of the $^{14}$N(p,$\gamma$)$^{15}$O cross section, CNO neutrinos constituted 1.6\% of the total flux from the sun in a standard solar model \cite{BP04}.  With the revised cross section, the fraction is 0.8\% \cite{BS05}.  An analysis \cite{Altmann:2005ix} by the GNO collaboration  using Ga data, the luminosity constraint, and the SSM value for the $^7$Be flux, gave as a central value for the CNO flux fraction 0.8\% (!), and a 3$\sigma$-limit of 6.5\%.

\subsection{$^7$Be(p,$\gamma$)$^8$B Reaction}

The cross section of this reaction at the Gamow peak, about 20 keV, is determined by means of theoretical extrapolation from higher energies at which laboratory measurements can be made.  The resulting accuracy has thus both a theoretical and an experimental component.  Recently, measurements of this cross section with an overall uncertainty of 3.8\% have been reported \cite{Junghans:2003bd}.   At this level of precision, this cross section plays a minor role in the overall uncertainty in the SSM prediction of the $^8$B flux.  Opacities and the cross section for $^3$He($\alpha,\gamma$)$^7$Be make substantially larger contributions. 

\subsection{$^3$He($\alpha,\gamma$)$^7$Be}

Two experimental methods for determining the cross section for this process in the laboratory, direct measurement of capture gammas, and the quantitative assay of the amount of $^7$Be produced, do not agree well, and as a result the rate of this reaction in the sun is uncertain by about 9\%.  Several laboratories are undertaking new measurements (LUNA, the University of Washington, and others). 

\subsection{Neutrino spectrum from $^8$B decay}

The shape of the neutrino spectrum from $^8$B decay enters directly into determinations of neutrino oscillation parameters, quite independent of any solar models.  The shape is not determined in the usual way from weak-interaction theory, because the final state in $^8$Be  is unbound and broad, with a beta strength function that calls for experimental measurement.   For some time the standard was an analysis \cite{Bahc6} of several experiments undertaken for reasons unrelated to solar neutrino research.  A new measurement by Ortiz \ea \ \cite{Ortiz:2000nf} indicated that the neutrino spectrum at high energies was significantly harder, and that spectrum was used in the analysis of SK and SNO data.   However, still more recent experimental work by Winter \ea \ \cite{Winter:2003ac,Winter:2004kf} gave results in agreement with the earlier spectra and in disagreement with Ortiz \ea.   Results obtained in a new experiment by Bhattacharya \ea \ \cite{adelb} are in excellent agreement with the results of Winter \ea.

\subsection{The {\em hep} reaction}

As described in more detail in Haxton \ea \ \cite{Haxton}, the {\em hep} reaction has a very low intensity and no significant role in energy production in the sun, but the very high endpoint energy (18.77 MeV) makes the reaction potentially observable.  Both SNO and SK are in the process of analyzing data to observe or set limits on this process. The s-wave capture is very hindered owing to different principal quantum numbers in the initial and final state wave functions, and about 40\% of the capture proceeds via the p-wave.  As a result the beta spectrum does not have a standard allowed shape, but no calculation of the spectral shape has yet been made.  The experimental groups need that calculation in order to extract a total rate from the high-energy events above the $^8$B endpoint. 

\subsection{Possible ground-state decay of $^8$B}

The decay of $^8$B directly to the ground state of $^8$Be involves a spin change of 2, and so is very hindered.  However, the high energy of the resulting neutrinos would place them in the same energy region as the observable part of the  {\em hep} reaction, and so it is important to determine if such a branch exists. Preliminary data taken at the University of Washington by Bacrania and Storm \cite{Bacrania} indicates the branch is less than 10$^{-4}$ and thus not a significant contributor. 

\section{Non-Standard-Model Scenarios}

A number of `new physics'  ideas beyond the now-standard LMA solution with 3 active species have been proposed that could be tested with present and future solar neutrino experimental data.  These include sterile neutrino admixtures \cite{Pulido:2005yc,Dev:2005px,deHolanda:2003tx,Bahcall:2002zh}, neutrino magnetic moments and resonant spin-flavor precession \cite{Friedland:2005xh}, and non-standard-model interactions \cite{Friedland:2004pp}.  

The analysis above, illustrated in Fig. \ref{fig:lum}, establishes the equality within experimental errors of the electromagnetic luminosity and the active-flavor neutrino luminosity.  As such, even at the present level of accuracy,  it sets a very general limit on the contribution of sterile neutrinos to the solar neutrino flux, but one that is also considerably more restrictive than those derived with reference to, for example, the SSM prediction of the $^8$B flux.  At the 90\% confidence level, sterile neutrinos do not constitute more than roughly 12\% of the flux. Consideration of the potential of future solar neutrino experiments suggests that highly sensitive tests will become possible.

\section{Future Solar Neutrino Experiments}

The SNO experiment will cease data-taking at the end of 2006, but SAGE and SK will continue.  They will, it is to be hoped, be joined by several new experiments aimed at precision investigation of the region below 5 MeV in the solar neutrino spectrum.  Two experiments, Borexino and KamLAND Solar, are approved and under construction.  Three others,  LENS, SNO+, and CLEAN, are in an active R\&D phase that will most likely be followed by  full proposals.  Two experiments, HERON and MOON Solar, have undergone extensive R\&D but are not at present on a trajectory to a full proposal.

\subsection{Borexino}

A liquid-scintillator ES experiment with a 100 Mg fiducial mass, Borexino is under construction in LNGS (Gran Sasso) \cite{Galbiati:2005gs}.    An accidental release of pseudocumene scintillator to the environment in 2003 precipitated a suspension of construction while remedies were put in place to prevent a similar occurrence in the future.  Most of that work has now been completed and it is expected that permission will be forthcoming to complete the detector as planned. 

\subsection{KamLAND Solar}

The KamLAND detector is already in operation as a reactor antineutrino detector.  The antineutrino signal provides a delayed coincidence tag (neutron capture) that completely rejects single-event backgrounds.  In order to make the detector suitable for detecting low-energy solar neutrinos by elastic scattering, reductions of contained $^{210}$Pb, $^{40}$K, $^{85}$Kr, $^{220}$Rn, $^{232}$Th, and $^{238}$U by factors ranging up to 10$^6$ are required.  A distillation plant is under construction.  Distillation removes the PPO fluor from the scintillator, and clean replacement material must be found and reintroduced.  Another difficulty is cosmogenic activity at the depth of Kamioka, about 2000 mwe.  Many such activities are sufficiently short lived that the interval following the passage of a muon can be vetoed, but 20-minute $^{11}$C requires special treatment.  It nevertheless leaves a clear window in which the $^7$Be solar neutrino line can be observed. 

When KamLAND Solar will be ready for data taking will depend on the performance of the distillation plant and other purification measures.  Tests will begin in 2006.  

\subsection{SNO+}

While the heavy water from SNO is to be returned to the owners in 2007, the potential use of the cavity, acrylic vessel, and phototube array has not gone unrecognized.  SNO+ is a proposal to fill the acrylic vessel with liquid scintillator \cite{Chen:2005yi}.  That would result in a detector of similar size and properties as KamLAND, but at the 6010-mwe depth of SNO.  Cosmogenic activities would not be a concern, and the reactor antineutrino ``background'' would also be considerably smaller.  One potentially thorny problem, the aggressive nature of most scintillator liquids to acrylic, has been resolved with the discovery \cite{Chen:2005yi} that linear alkylbenzenes,  common chemical intermediates in detergent manufacturing, make an efficient and non-aggressive scintillator. 

\subsection{CLEAN and HERON}

Noble liquids are highly efficient scintillators transparent to their own radiation, and among them He and Ne possess no long-lived isotopes that represent backgrounds to solar neutrinos.  Their low boiling points permit fractionation to remove other troublesome contaminants such as $^{85}$Kr.  The CLEAN concept \cite{McKinsey:2004rk} involves LNe and a wavelength-shifter on the surface of the container where phototubes are mounted.  Event positions can be reconstructed via the luminance distribution at the surface.  HERON \cite{Lanou:2005ku} makes use of both scintillation light and electrons liberated by ionizing radiation to reconstruct the energy and position within the fiducial volume, and to reject events that have multiple interaction sites, such as Compton interactions.  

\subsection{LENS and MOON}

The LENS experiment, an active, spectroscopic, charged-current experiment, has been under development in various forms for many years \cite{Raghavan:1980pi}.  The basic principle is to make use of a neutrino capture to an excited state, the decay of which leads to an isomer.  The delayed decay of the isomer provides a tag to reject backgrounds.  The most favorable case once again appears to be $^{115}$In, with a 4.76-$\mu$s isomer in the daughter $^{115}$Sn.  The Q-value to this state is only -114 keV, but unfortunately, by the standards of solar neutrino rates, In is intensely radioactive, with a half-life of $6 \times 10^{14}$ years.  Thus, most R\&D effort has gone into strategies for mitigating this inherent background.   The 498-keV beta background and associated bremsstrahlung  intrude as chance coincidences.  Recent successes in development of an In-loaded scintillator have produced 8\% loading with attenuation scarcely worse than pure pseudocumene, good resolution, and long-term stablility \cite{Raghavan:2005Taup}.  Good resolution helps reject the continuous beta background.  Another advance has been development of an array of cubical cells to contain the scintillator and guide light predominantly to phototubes on the cell's major axes.  In this way good segmentation can be obtained without excessive numbers of channels.   A detector of 125-190 Mg fiducial mass, containing 10-15 Mg of In, is currently envisaged.

The MOON concept \cite{Hazama:2005kv,MOON} for solar neutrinos takes advantage of the very large matrix element for $\nu_e$ capture connecting the ground state of $^{100}$Mo to the $1^+$ ground state of $^{100}$Tc, with a Q-value of -168 keV.  The subsequent beta decay of 16-s $^{100}$Tc provides a tag to identify the solar neutrino capture and reject backgrounds.  The primary technical challenge is the $2\nu\beta\beta$  background from the decay of $10^{18}$-y $^{100}$Mo.  In order to reduce the random coincidences to a level below the signal, the detector must be effectively subdivided with a volume resolution corresponding to less than 100 mg of $^{100}$Mo.

\section{Conclusions}

Solar neutrino research has a luminous past and a bright future.  It has contributed in a major way to a revolution in fundamental physics that has required the first revision to the Standard Model of particles and fields.  Neutrinos are strongly mixed in flavor and have non-zero masses.  The next steps for the field are generally agreed upon: precise, spectroscopic measurement of the low-energy fluxes from the sun.  A direct confrontation of the luminosity constraint is a matter of considerable importance even beyond the fields of physics and astrophysics.  The combination of intensity and remoteness of the solar neutrino source gives it power to test for the presence of non-standard neutrino physics inaccessible to other kinds of experiment.   The unexpected can continue to be expected.

\end{document}